\documentclass[showpacs,aps,prl,amsmath,amssymb,twocolumn]{revtex4}
\usepackage{CJK}
\newcommand{\dbar} {\ensuremath{\,\mathchar'26\mkern-12mu d}}
\usepackage[latin1]{inputenc}
\usepackage{graphicx}
\usepackage{bm}
\usepackage{xcolor}
\usepackage{array}
\usepackage{braket}
\usepackage{mathtools}
\usepackage{microtype}
\usepackage{wasysym}
\renewcommand{\baselinestretch}{1.0}
\begin{document}

\renewcommand{\baselinestretch}{1.0}
\title{Relating Heat and Entanglement in Strong Coupling Thermodynamics}

\author{Bert\'{u}lio de Lima Bernardo$^{1,2}$}

\affiliation{$^{1}$Departamento de F\'{\i}sica, Universidade Federal da Para\'{\i}ba, 58051-900 Jo\~ao Pessoa, PB, Brazil\\
$^{2}$Departamento de F\'{\i}sica, Universidade Federal de Campina Grande, Caixa Postal 10071, 58109-970 Campina Grande-PB, Brazil}

\email{bertulio.fisica@gmail.com}

\begin{abstract}

Explaining the influence of strong coupling in the dynamics of open quantum systems is one of the most challenging issues in the rapidly growing field of quantum thermodynamics. By using a particular definition of heat, we develop a new approach to study thermodynamics in the strong coupling regime, which takes into account quantum resources as coherence and entanglement. We apply the method to calculate the time-dependent thermodynamic properties of a system and an environment interacting via the generalized amplitude-damping channel (GADC). The results indicate that the transient imbalance between heat dissipated and heat absorbed that occurs in the process is responsible for the generation of system-environment entanglement.                                                
   
\end{abstract}

\maketitle


{\it Introduction}.--- Classical thermodynamics intrinsically relies on the assumption that the system under analysis is weakly
coupled to its surroundings. This is because the energy of the interacting elements of the body's surface is negligible compared to the energy of the bulk. In this regime, one can always treat the states and the energetic properties of the system and environment separately, which allows us, for example, to equate the energy dissipated by the system with the heat absorbed by the reservoir \cite{goold,deffner,binder,jarzynski,hsiang}. On the other hand, this weak coupling limit is not in general justified for quantum open systems, since the system-environment interaction involves a large fraction of the system's constituents \cite{breuer,talkner}. In this strong coupling regime, stochastic and quantum effects become important \cite{kockum}, and the usual approach to this problem begins with the decomposition of the total system-environment Hamiltonian in the form 
\begin{equation}
\label{hamiltonain}
\hat{H} = \hat{H}_{\mathcal{S}} + \hat{H}_{\mathcal{E}} + \hat{H}_{\mathcal{SE}},
\end{equation}  where the operators $\hat{H}_{\mathcal{S}}$ and $\hat{H}_{\mathcal{E}}$ are the bare Hamiltonians of the system and environment, respectively, and $\hat{H}_{\mathcal{SE}}$ is the interaction Hamiltonian, which cannot be neglected.

Some particularly important problems are often pointed out when the coupling is strong. One is how to partition the internal energy into work and heat, as dictated by the first law, $\Delta U = W +Q$. In this case, the notion of work is less problematic because, as usual, it only
depends on system variables, but the definition of heat has been shown to be more difficult and ambiguous \cite{talkner,seifert,rivas}. Another question is that the reduced density matrix of the open system is not supposed to contain all the information necessary to describe the thermodynamic properties; instead, information originating from the system-environment interaction should be included \cite{solinas,schmidt,gogolin,subasi,ness,strasberg}. In this article, we present a framework to study the thermodynamics of open quantum systems in the strong coupling regime, based on a recently proposed quantum version of the first law of thermodynamics \cite{bert2}. The formalism circumvents the limitations indicated above, in the sense that an unambiguous definition of heat is provided, which considers quantum effects as coherence and entanglement, and the reduced density matrices of the subsystems involved are capable of providing full information about their average thermodynamic behavior. To illustrate the practical implications, we use the method to investigate the thermodynamics of the generalized amplitude-damping channel (GADC). Interestingly, we find that the system-bath entanglement in this case is generated at the cost of the heat asymmetry that naturally emerges during the interaction process.

\section{First law of quantum thermodynamics}

In this section we review the quantum version of the first law of thermodynamics put forward by the author in Ref. \cite{bert2}. Consider a generic quantum, physical system operating as a working substance, whose Hamiltonian reads $\hat{H} = \sum_{n} E_{n} \ket{n}\bra{n}$, with the $n$-th energy eigenvalue and eigenstate given by $E_{n} = \braket{n|\hat{H}|n}$ and $\ket{n}$, respectively. We also define the density operator of the system as
$\hat{\rho} = \sum_{k} \rho_{k} \ket{k}\bra{k}$, where $\rho_{k} = \braket{k|\hat{\rho}|k}$ and $\ket{k}$ are the eigenvalues and eigenkets, respectively. From a statistical perspective, we can define the internal energy of this system as the average of $\hat{H}$, i.e., $U = \langle \hat{H} \rangle =$ tr$\{\hat{\rho} \hat{H} \}$. However, since the trace operation is basis-independent, we can calculate $U$ using either the eigenbasis $\{ \ket{n} \}$ of $\hat{H}$ or the eigenbasis $\{ \ket{k} \}$ of $\hat{\rho}$. In the first case, which we label as C1, we have that $U = \sum_{n} P_{n} E_{n}$, where $P_{n} = \braket{n|\hat{\rho}|n}$ is the occupation probability of the $n$-th energy level, while in the second case, which we label as C2, one finds $U = \sum_{k} \rho_{k} \epsilon_{k}$, with $\epsilon_{k} = \braket{k|\hat{H}|k}$ as the diagonal elements of $\hat{H}$ represented in the $\{ \ket{k} \}$ basis. 

The result of the internal energy obtained in the case C1 allows us to write $dU = \sum_{n} [E_{n} d P_{n} + P_{n} d E_{n}]$. Now, we are in a position to define the work done on the system in an infinitesimal quantum process, $\dbar W$. In doing so, we first recall the classical concept of work: ``the work realized on or by the working substance is the change in the internal energy produced by modifications in the generalized coordinates'' \cite{kittel,callen,landau}. In a quantum-mechanical setting, a change in the generalized coordinates of the system, which may include the volume, external electric and magnetic fields, or the gravitational potential, leads naturally to a modification in the energy spectrum $E_{n}$, as for example in quantum dynamics satisfying the adiabatic approximation \cite{born,kato,messiah}, or in the so-called shortcuts to adiabaticity \cite{torr,odelin,adc,deng,bert}. Thus, from the expression of $dU$ above, we can identify the infinitesimal work as $\dbar W \coloneqq \sum_{n} P_{n} d E_{n}$.

Let us now study the case C2. In this scenario, we have that $dU = \sum_{k} [\epsilon_{k} d\rho_{k} + \rho_{k} d\epsilon_{k}]$, which permits us to define the heat exchange by the working substance in an infinitesimal quantum process, $\dbar Q$. In classical thermodynamics, the concept of heat reads: ``the heat exchanged between the working substance and the external environment corresponds to the change in the internal energy that is accompanied by entropy change'' \cite{kittel,landau}. In order to generalize this concept to the quantum realm, we first recall that the von Neumann entropy of the system is given by $S(\hat{\rho})=-$tr$\{ \hat{\rho} \log \hat{\rho}\} = - \sum_{k} \rho_{k} \log \rho_{k}$. Then, for an infinitesimal trace-preserving quantum transformation, we have that $dS = - \sum_{k} [\log (\rho_{k}) d\rho_{k}]$, where we used the fact that $\sum_{k} d \rho_{k} = 0$, because $\sum_{k} \rho_{k} = 1$. From the results of $dS$ and $dU$ obtained in this case C2, if we invoke the above classical definition of heat, we can identify the quantum heat as $\dbar Q \coloneqq \sum_{k} \epsilon_{k} d \rho_{k}$. Indeed, this is the only part of $dU$ which accompanies entropy change. 

Having the definitions of $\dbar W$ and $\dbar Q$ above, it can be easily verified that $dU \neq \dbar W + \dbar Q$, in apparent contradiction with the first law of thermodynamics. However, as shown in Ref. \cite{bert2}, the missing energetic contribution is given by the infinitesimal quantity $\dbar \mathcal{C} = \sum_{n}\sum_{k} (E_{n} \rho_{k}) d \left[ |c_{n,k}|^{2} \right]$, where $c_{n,k} = \braket{n|k}$. In fact, it can be demonstrated that $dU = \dbar W + \dbar Q + \dbar \mathcal{C}$. Observe that the contribution of $\dbar \mathcal{C}$ only exists if the thermodynamic process in question involves some change in the quantum coherence of the system in the energy basis, i.e., when the coefficients $|c_{n,k}|$ are time-dependent. For this reason, we will refer to the quantity $\mathcal{C}$ here as {\it coherent energy}. Of course, this is only relevant in specific quantum processes, and the usual form of the first law, $dU = \dbar W + \dbar Q$, is recovered in the classical limit. 

Overall, the function $\mathcal{C}$, which is not compatible with the classical definitions of work and heat, expresses the energetic contribution of the dynamics of coherence in the first law. This is why we give it an independent classification. For finite-time processes, the work, heat and coherent energy can be found by integration of the respective differentials (see details in Ref. \cite{bert2}):  
\begin{equation}
\label{1}
W(t)  =  \sum_{n} \sum_{k} \int_{0}^{t}  \rho_{k} |c_{n,k}|^{2} \frac{d E_{n}}{dt'} dt',
\end{equation}
\begin{equation}
\label{2}
Q(t)  =  \sum_{n} \sum_{k} \int_{0}^{t}  E_{n} |c_{n,k}|^{2} \frac{d \rho_{k}}{dt'} dt',
\end{equation}    
\begin{align}
\label{3}
\mathcal{C}(t)
&= \sum_{n}\sum_{k} \int_{0}^{t} (E_{n} \rho_{k}) \frac{d}{dt'} |c_{n,k}|^{2} dt'.
\end{align}
The change in the internal energy is given by $\Delta U(t) = W(t) + Q(t) + \mathcal{C}(t)$. In what follows, we apply these results to study the energy exchanges that occur in a particular strong system-environment interaction. 

\section{Physical model}

In order to study the evolution of an open quantum system $\mathcal{S}$, one usually considers it as part of a larger closed system, which also includes the environment $\mathcal{E}$, undergoing a unitary transformation $\hat{\mathcal{U}}$ that depends on the total Hamiltonian as that of Eq. (\ref{hamiltonain}). Following this reasoning, we illustrate our description of strong coupling thermodynamics by examining a model based on the generalized amplitude-damping channel (GADC) \cite{khatri}. This model is a useful tool to describe the dynamics of a qubit system in contact with a thermal bath with finite temperature. For instance, the GADC has been used to characterize a spin-1/2 system coupled to an interacting spin chain at nonzero temperature \cite{bose, goold2}, the influence of  noise in superconducting-circuit-based quantum computing \cite{chirolli}, and the finite-temperature thermal noise in linear optical systems \cite{zou}. Here, we consider a particular approach consisting of an open two-level atomic system $\mathcal{S}$ interacting with two levels of a finite environment $\mathcal{E}$ initially in a thermal state, as depicted
in Fig. 1. 

\begin{figure}[ht]
\centerline{\includegraphics[width=6.0cm]{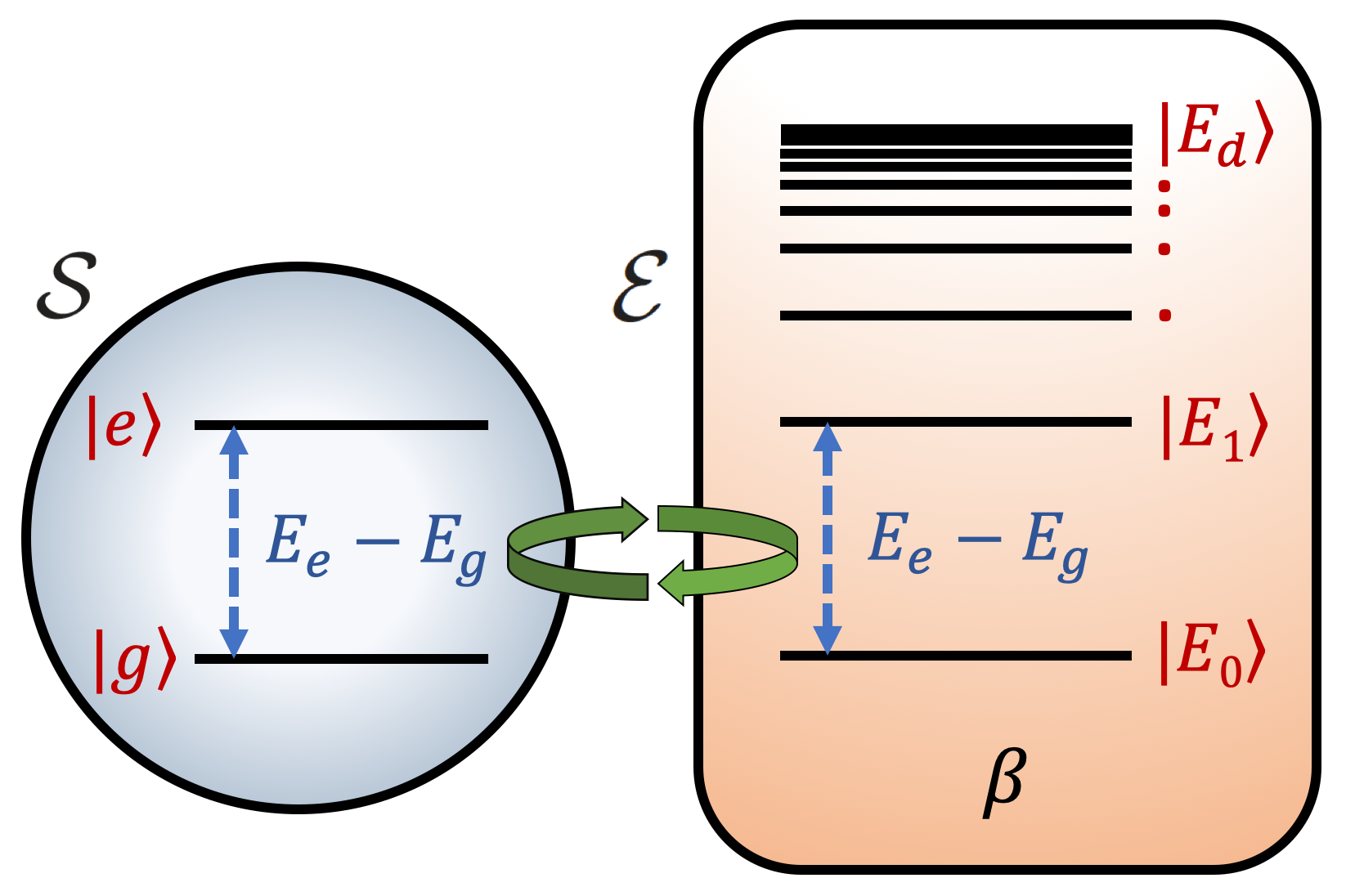}}
\caption{Schematic representing a two-level atomic system $\mathcal{S}$ interacting with an environment $\mathcal{E}$, which is initially in a thermal state at inverse temperature $\beta$.}
\label{setup}
\end{figure}

The ground and excited states of the system, $\ket{g}$ and $\ket{e}$, have energies $E_{g}$ and $E_{e}$, respectively, so that the free Hamiltonian of the system is given by $\hat{H}_{\mathcal{S}} = E_{g}\ket{g}\bra{g} + E_{e}\ket{e}\bra{e}$. On the other hand, we assume that the environment $\mathcal{E}$ has a small number of states, $d + 1$, and that only the transitions between the states $\ket{E_{0}}$ and $\ket{E_{1}}$ are capable of causing some influence in $\mathcal{S}$. The free environmental Hamiltonian and initial thermal state are given by $\hat{H}_{\mathcal{E}} = \sum_{i=0}^{d} E_{i} \ket{E_{i}}\bra{E_{i}}$ and $\hat{\rho}_{\mathcal{E}} (0)= e^{-\beta \hat{H}_{\mathcal{E}}}/ \mathcal{Z}_{\mathcal{E}}$, respectively, where $\mathcal{Z}_{\mathcal{E}} =$ tr$[e^{-\beta \hat{H}_{\mathcal{E}}}]$ is the partition function, and $\beta$ the inverse temperature. As a matter of fact, since we are assuming that $\mathcal{S}$ interacts with $\mathcal{E}$ only via transitions that occur between $\ket{E_{0}}$ and $\ket{E_{1}}$, the environment can be effectively represented as a qubit (which we call $E$), with initial state $\hat{\rho}_{E}(0) = w_{0}\ket{E_{0}}\bra{E_{0}} + w_{1}\ket{E_{1}}\bra{E_{1}}$, where the probabilistic weights obey $w_{1} = w_{0} e^{-\beta(E_{1} - E_{0})}$ and $w_{0} + w_{1} = 1$. We also see from Fig. 1 that $E_{1} - E_{0} = E_{e} - E_{g}$.

To describe the system-environment dynamics, we use the joint $\mathcal{SE}$ basis $\{ \ket{g,E_{0}},\ket{g,E_{1}},\ket{e,E_{0}},\ket{e,E_{1}} \}$ and establish a (probabilistic) unitary evolution such that 
\begin{equation}
\label{4}
\ket{g,E_{0}} \rightarrow \ket{g,E_{0}},  
\end{equation}
\begin{equation}
\label{5}
\ket{g,E_{1}} \rightarrow \sqrt{1-p} \ket{g,E_{1}} + \sqrt{p} \ket{e,E_{0}},
\end{equation}
\begin{equation}
\label{6}
\ket{e,E_{0}} \rightarrow \sqrt{1-p} \ket{e,E_{0}} + \sqrt{p} \ket{g,E_{1}}, \end{equation}
\begin{equation}
\label{7}
\ket{e,E_{1}} \rightarrow \ket{e,E_{1}}.  
\end{equation}
These interaction rules can be interpreted as follows: i) Eq. (\ref{4}) says that if $\mathcal{S}$ starts out in the ground state and $\mathcal{E}$ has no excitation (e.g., zero temperature), no transition occurs; ii) Eq. (\ref{5}) indicates that if $\mathcal{S}$ is in the ground state and $\mathcal{E}$ in the first excited state, after a given time interval $\tau$, there is a probability $p$ that $\mathcal{S}$ becomes excited and $\mathcal{E}$ decays to the fundamental state; iii) Eq. (\ref{6}) says that if $\mathcal{S}$ is in the excited state and $\mathcal{E}$ in the fundamental one, after the time $\tau$, $\mathcal{S}$ decays to the ground state with probability $p$, and $\mathcal{E}$ is led to the first excited state; iv) Eq. (\ref{7}) tells us that, if $\mathcal{E}$ is in the first excited state, the state $\ket{e}$ of $\mathcal{S}$ has a comparatively longer lifetime, so that no transition is expected during the time $\tau$. The longer lifetime of $\ket{e,E_{1}}$ in comparison with $\ket{e,E_{0}}$ can be justified by the fact that, in some strong coupling cases, the metastability of quantum states is sensitive to the environmental conditions \cite{maci, boite,valenti}.     

According to Eqs. (\ref{4}) to (\ref{7}), the matrix representation of the evolution of $\mathcal{S}E$ is given by
\begin{equation}
\label{8}
\hat{\mathcal{U}} = 
\begin{bmatrix}
    1 & 0 & 0 & 0 \\
    0 & \sqrt{1-p} & \sqrt{p} &  0 \\
    0 & \sqrt{p} & \sqrt{1-p} &  0\\
    0 & 0 & 0 & 1
\end{bmatrix},
\end{equation}
with $p \in [0,1]$. Note that, in the limit where $p=1$, $\hat{\mathcal{U}}$ reduces to the SWAP gate \cite{nielsen}. Now, if we assume that $\mathcal{S}$ and $E$ are initially uncorrelated, $\hat{\rho}_{\mathcal{S}E}(0) = \hat{\rho}_{\mathcal{S}}(0) \otimes \hat{\rho}_{E}(0)$, we can describe the evolution of $\mathcal{S}$ through the quantum channel $\hat{\rho}_{\mathcal{S}}(0) \rightarrow \Phi[\hat{\rho}_{\mathcal{S}}(0)]=$ tr$_{E}[\hat{\mathcal{U}} ( \hat{\rho}_{\mathcal{S}}(0) \otimes \hat{\rho}_{E}(0) ) \hat{\mathcal{U}}^{\dagger}]$, where tr$_{E}$ denotes trace over the environment states $\ket{E_{0}}$ and $\ket{E_{1}}$. It can be shown that $\Phi[\hat{\rho}_{\mathcal{S}}(0)] = \sum_{ij=0,1} \hat{K}_{ij} \hat{\rho}_{\mathcal{S}}(0) \hat{K}^{\dagger}_{ij}$ is a completely positive trace preserving (CPTP) map with Kraus operators $\hat{K}_{ij} = \sqrt{w_{i}} \braket{E_{j}|\hat{U}|E_{i}}$ given by: $\hat{K}_{00} = \sqrt{w_{0}} (\ket{g}\bra{g} + \sqrt{1-p}\ket{e}\bra{e})$, $\hat{K}_{01} = \sqrt{w_{0}} ( \sqrt{p}\ket{g}\bra{e})$, $\hat{K}_{10} = \sqrt{w_{1}} ( \sqrt{p}\ket{e}\bra{g})$, and $\hat{K}_{11} = \sqrt{w_{1}} (\sqrt{1-p} \ket{g}\bra{g} + \ket{e}\bra{e})$, which satisfy $\sum_{ij=0,1} \hat{K}^{\dagger}_{ij} \hat{K}_{ij} = \mathbb{I}_{\mathcal{S}}$ \cite{nielsen,preskill}. This is the GADC \cite{khatri}.    

We next turn to investigating the dynamics of $E$. This is dictated by the channel $\hat{\rho}_{E}(0) \rightarrow \Lambda[\hat{\rho}_{E}(0)]=$ tr$_{\mathcal{S}}[\hat{\mathcal{U}} ( \hat{\rho}_{\mathcal{S}}(0) \otimes \hat{\rho}_{E}(0) ) \hat{\mathcal{U}}^{\dagger}] $, where tr$_{\mathcal{S}}$ denotes trace over the system states, $\ket{g}$ and $\ket{e}$, which yields $\Lambda[\hat{\rho}_{E}(0)] = \sum_{k=0,1} \hat{L}_{k} \hat{\rho}_{E}(0) \hat{L}^{\dagger}_{k}$. For simplicity, we assume that the system is prepared in the pure state $\ket{\psi(0)} = \alpha \ket{g} + \sqrt{1 - \alpha^2} \ket{e}$, with $\alpha \in \mathbb{R}$ (this assumption does not invalidate the generality of the results, i.e., a general mixed state could be equally used). In this case, the two Kraus operators are given by $\hat{L}_{0} = \braket{g|\hat{U}|\psi(0)} = \alpha \ket{E_{0}}\bra{E_{0}} + \sqrt{p(1-\alpha^2)} \ket{E_{1}}\bra{E_{0}} + \sqrt{1-p}\alpha \ket{E_{1}}\bra{E_{1}}$, and $\hat{L}_{1} = \braket{e|\hat{U}|\psi(0)} = \sqrt{(1-p)(1-\alpha^2)} \ket{E_{0}}\bra{E_{0}} + \sqrt{p}\alpha \ket{E_{0}}\bra{E_{1}} + \sqrt{1-\alpha^2} \ket{E_{1}}\bra{E_{1}}$. This is also a CPTP map, in which $\sum_{k=0,1} \hat{L}^{\dagger}_{k} \hat{L}_{k} = \mathbb{I}_{E}$.  

As a result of the application of the maps, the states of $\mathcal{S}$ and $E$ become
\begin{align}
\label{9.1}
 \Phi [\hat{\rho}_{\mathcal{S}}(0)] =
\begin{bmatrix}
    A_{11} & A_{12} \\
    A_{21} & A_{22}\\
\end{bmatrix},
&&
\Lambda [\hat{\rho}_{E}(0)] = 
\begin{bmatrix}
    B_{11} & B_{12} \\
    B_{21} & B_{22}\\
\end{bmatrix},
\end{align}
respectively. The entries of $\Phi [\hat{\rho}_{\mathcal{S}}(0)]$ are given by $A_{11} = [\alpha^2 + (1 - \alpha^2)p]w_{0} + \alpha^2 (1-p) w_{1}$, $A_{12} = A_{21} = \alpha \sqrt{(1 - \alpha^2) (1-p)} $, and $A_{22} = (1 - \alpha^2)(1-p) w_{0} + [(1-\alpha^2) + \alpha^2 p] w_{1}$, whereas the entries of $\Lambda [\hat{\rho}_{E}(0)]$ are $B_{11} = [\alpha^2 + (1 - a^2)(1-p)]w_{0} + \alpha^2 p w_{1}$, $B_{12} = B_{21} = \alpha \sqrt{(1 - \alpha^2)p} $, and $B_{22} = (1 - \alpha^2)p w_{0} + [(1-\alpha^2) + \alpha^2 (1-p)] w_{1}$. It can be seen that, although the unitary (reversible) evolution of $\mathcal{S}E$, the partial trace operations used to construct the above maps lead to non-unitary (irreversible) evolution of $\mathcal{S}$ and $E$, individually.

To express the time evolution of $\mathcal{S}$ and $E$, we will assume that the probability of a quantum transition event per unit time is $\Gamma$, in such a way that $p = \Gamma \Delta t \ll 1$ for a short time interval $\Delta t$. Then, the evolution of the system and the environment after a time $t = n \Delta t$ is a result of the application of the respective maps $n$ times in succession. This assumption is equivalent to assuming that the evolution of $\mathcal{S}$ and $E$ are {\it Markovian}, i.e., the influence of the quantum channels acting on $\mathcal{S}$ and $E$ are completely determined by the respective quantum states at each time step \cite{breuer2,vega,li}. Accordingly, these transformations can be implemented as $\hat{\rho}_{\mathcal{S}}(t) = \Phi^{n} [\hat{\rho}_{\mathcal{S}}(0)]$ and $\hat{\rho}_{E}(t) = \Lambda^{n} [\hat{\rho}_{E}(0)]$, which make the probabilistic factors change based on the rule $(1-p) \rightarrow (1-p)^{n} = \lim_{n \to \infty} \left(1-\frac{\Gamma t}{n} \right)^n = e
^{- \Gamma t}$, where we assumed $\Delta t \rightarrow 0$ \cite{preskill}. In this form, we can write 
\begin{align}
\label{10}
\hat{\rho}_{\mathcal{S}}(t)  = 
\begin{bmatrix}
    A_{11}(t) & A_{12}(t) \\
    A_{21}(t) & A_{22}(t)\\
\end{bmatrix},
&&
\hat{\rho}_{E}(t) = 
\begin{bmatrix}
    B_{11}(t) & B_{12}(t) \\
    B_{21}(t) & B_{22}(t)\\
\end{bmatrix}.
\end{align}

In this case, the entries of $\hat{\rho}_{\mathcal{S}}(t)$ are given by $A_{11}(t) = [\alpha^2 + (1 - \alpha^2)\delta(t)]w_{0} + \alpha^2 \gamma (t) w_{1}$, $A_{12}(t) = A_{21}(t) = \alpha \sqrt{1 - \alpha^2} [\gamma (t)]^{1/2}$, and $A_{22}(t) = (1 - \alpha^2)\gamma(t) w_{0} + [(1-\alpha^2) + \alpha^2 \delta(t)] w_{1}$. In turn, the entries of $\hat{\rho}_{E}(t)$ are $B_{11}(t) = [\alpha^2 + (1 - a^2)\gamma(t)]w_{0} + \alpha^2 \delta (t) w_{1}$, $B_{12}(t) = B_{21}(t) = \alpha \sqrt{1 - \alpha^2} [\delta (t)]^{1/2}$, and $B_{22}(t) = (1 - \alpha^2)\delta(t) w_{0} + [(1-\alpha^2) + \alpha^2 \gamma(t)] w_{1}$. In these equations we have that $\gamma (t) = e^{-\Gamma t}$ and $\delta (t) = 1- e^{-\Gamma t}$. The consistency of the states in Eq. (\ref{10}) in this physical scenario relies on the tacit assumption that the characteristic time scale
of the system-environment interaction, $\tau_{int} = 1/ \Gamma$, is much faster than the characteristic time of the other transitions involving the environment states $\ket{E_{0}}$ and $\ket{E_{1}}$. It is worth mentioning that the positive constant $\Gamma$, which can also be interpreted as a decoherence rate of $\mathcal{S}$, is a characteristic of the Markovian dynamics assumed here \cite{breuer2,vega,li}. As a consequence, there is a flow of information from the system to the environment before they reach a steady state. As we shall see, this behavior is indicated in the graphs of the thermodynamic quantities shown in Fig. 2.

\section {Thermodynamics of the model}

Having found the density operators of the system $\mathcal{S}$ and the environment qubit $E$, $\hat{\rho}_{\mathcal{S}}(t)$ and $\hat{\rho}_{E}(t)$, and considering the respective Hamiltonians, $\hat{H}_{\mathcal{S}} = E_{g}\ket{g}\bra{g} + E_{e}\ket{e}\bra{e}$ and $\hat{H}_{E} = E_{0} \ket{E_{0}}\bra{E_{0}} + E_{1} \ket{E_{1}}\bra{E_{1}}$, we can calculate the eigenvalues and eigenstates of these four operators. This information allows us to calculate the thermodynamic properties of $\mathcal{S}$ and $E$ as a function of time during the interaction process, according to Eqs. (\ref{1}) to (\ref{3}). As the energy eigenvalues of $\mathcal{S}$ and $E$ are time-independent, it is straightforward to see from Eq. (\ref{1}) that no work is done on the system and the environment, i.e., $W_{\mathcal{S}}(t) = W_{E}(t) = 0$. However, in order to calculate the heat exchange and the coherent energy of $\mathcal{S}$ and $E$, we first need to fix some parameters. As can be seen from the off-diagonal elements of $\hat{\rho}_{\mathcal{S}}(t)$ and $\hat{\rho}_{E}(t)$, the quantum coherence of these states varies only if $\alpha \neq 0,1$ \cite{strel,SM}. Thus, an interesting case to study is when $\alpha = 1/\sqrt{2}$. Let us also consider that the inverse temperature of the environment is $\beta = 1/ (E_{e} - E_{g})$. In this form, we have that $w_{0} \approx 0.73$ and $w_{1} \approx 0.27$. 

According to Eqs. (\ref{2}) and (\ref{3}), we are now able to calculate the heat exchange and coherent energy of the system, $Q_{\mathcal{S}}(t)$ and $\mathcal{C}_{\mathcal{S}}(t)$, and the environment, $Q_{E}(t)$ and $\mathcal{C}_{E}(t)$, as the GADC proceeds. 
Although these quantities can be computed analytically, the expressions are too cumbersome to be shown here. Instead, we plot the results as a function of time, as shown in Fig. 2.
It can be verified that
$\Delta U_{\mathcal{S}}(t) = - \Delta U_{E}(t) =$ tr$\{ \hat{H}_{\mathcal{S}}(\hat{\rho}_{\mathcal{S}}(t) - \hat{\rho}_{\mathcal{S}}(0))\}$, $\forall t$. However, we call attention
to the difference between $Q_{\mathcal{S}}(t)$ and $- Q_{E}(t)$, and between $\mathcal{C}_{\mathcal{S}}(t)$ and $- \mathcal{C}_{E}(t)$, especially for $t<4$. This result confirms that the weak-coupling condition, in which the relation $Q_{\mathcal{S}}(t) = - Q_{E}(t)$ is applicable, is not fulfilled in the present model. In order to quantify the discrepancy between the heat released by $E$ and the heat absorbed by $\mathcal {S}$, we introduce the quantity $Q_{\mathcal {S}E} (t) = |Q_{\mathcal {S}}(t) + Q_{E}(t)|$ that we call {\it heat asymmetry}, whose time-dependence is shown in Fig. 3. Note that, since $\Delta U_{\mathcal{S}}(t) = - \Delta U_{E}(t)$ and $W_{\mathcal{S}}(t) = W_{E}(t) = 0$, the event $Q_{\mathcal {S}E} (t) \neq 0$ is a consequence of the difference between the rates of entropy change of the system and the environment, i.e., $d S(\hat{\rho}_{\mathcal{S}})/dt \neq - d S(\hat{\rho}_{E})/dt$ \cite{SM}. Meantime, the quantity $Q_{\mathcal {S}E}(t)$ approaches zero for long times. 

\begin{figure}[ht]
\centerline{\includegraphics[width=6.5cm]{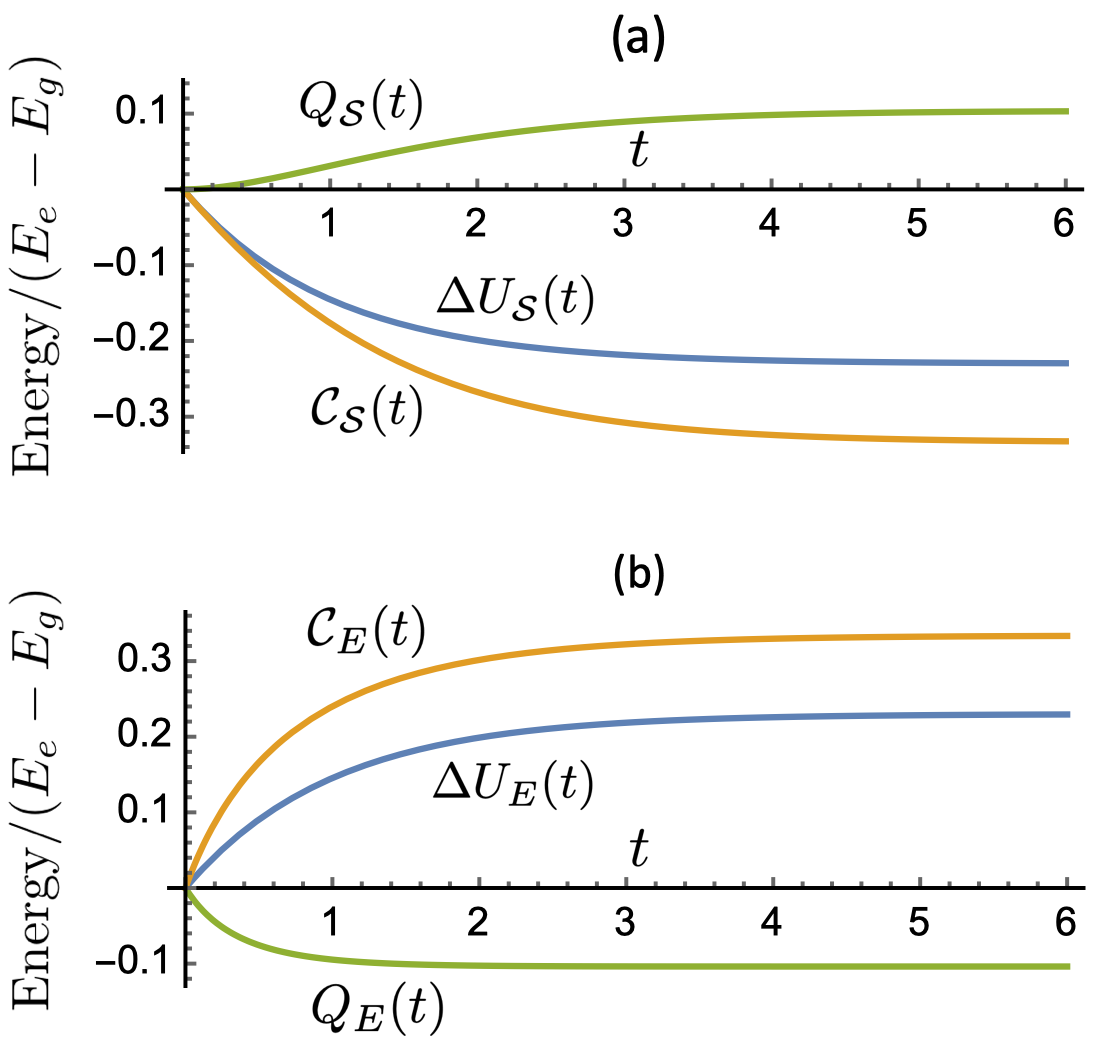}}
\caption{Time evolution of the heat exchange, coherent energy and internal energy of the (a) system $\mathcal{S}$ and (b) environment qubit $E$, interacting via the GADC. Initially, $\mathcal{S}$ is in a maximally coherent state, $\ket{+} = 1/ \sqrt{2} (\ket{g}+\ket{e})$, and $E$ is at thermal equilibrium with inverse temperature $\beta = 1/ (E_{e} - E_{g}) $. For simplicity, we assumed $\Gamma = 1$ in both panels.} \end{figure}

We also see from Eq.~(\ref{3}) that the coherent energy $\mathcal{C}(t)$ can be physically interpreted as the energy transfer to or from a system accompanied by coherence change. Based on this interpretation, we can say that the amount of coherent energy entailed in a quantum process depends both on the initial coherence of the system and, of course, the nature of the interaction with the environment. In general, quantum transformations that involve large amounts of coherent energy are those in which the system is initially in a high-coherence state, and the coupling with the environment causes strong dissipation and decoherence \cite{zurek,zurek2,schloss}. In this context, we observe that the coherent energy flow may become more prominent when the system is coupled to low-temperature environments, as in the spin-spin model \cite{prokof,dube}, and high-temperature environments, as in the Caldeira-Leggett model \cite{caldeira,weiss}. Actually, we note that coherent energy, as well as heat exchange, is not a quantity that is determined by the coupling strength.

\section{Quantum correlations}

We next focus on the study of the entanglement created due to the coupling between $\mathcal{S}$ and $E$. The quantification of entanglement for bipartite mixed states is not a trivial task. Nevertheless, since we reduced our problem to a qubit-qubit interaction, we can use the concept of negativity to characterize the system-environment entanglement. The negativity is an entanglement monotone given by \cite{vidal,horodecki}
\begin{equation}
\label{11}
\mathcal{N}(\hat{\rho}_{\mathcal{S}E}) =\frac{ \| \hat{\rho}_{\mathcal{S}E}^{T_{\mathcal{S}}} \| - 1}{2},
\end{equation}
where $\hat{\rho}_{\mathcal{S}E}$ denotes the density matrix of the composite system, comprising $\mathcal{S}$ and $E$, and $\hat{\rho}_{\mathcal{S}E}^{T_{\mathcal{S}}}$ the partial transpose of $\hat{\rho}_{\mathcal{S}E}$ with respect to the system. The trace norm of an operator $\hat{O}$ is defined as $\| \hat{O} \| = $ tr$\{ \sqrt{\hat{O} \hat{O}^{\dagger}} \}$. The negativity is also given by the sum of the absolute values of the negative eigenvalues of $\hat{\rho}_{\mathcal{S}E}^{T_{\mathcal{S}}}$, which vanishes for unentangled states. Previous studies have used the negativity to quantify system-environment quantum correlations \cite{hilt,wybo}. The time evolution of the operator $\hat{\rho}_{\mathcal{S}E}$ can be calculated from the initial uncorrelated state $\hat{\rho}_{\mathcal{S}E}(0) = \hat{\rho}_{\mathcal{S}}(0) \otimes \hat{\rho}_{E}(0)$ as $\hat{\rho}_{\mathcal{S}E}(t) = \hat{\mathcal{\mathcal{U}}} \hat{\rho}_{\mathcal{S}E}(0)  \hat{\mathcal{U}}^{\dagger}$, with $\hat{\mathcal{U}}$ given in Eq. (\ref{8}), along with the probabilistic rules considered here, $p \rightarrow 1 -e
^{- \Gamma t}$. This, in combination with Eq. (\ref{11}), permits us to calculate the system-environment entanglement as a function of time, $\mathcal{N}[\hat{\rho}_{\mathcal{S}E}(t)]$, for the GADC \cite{SM}. The result is displayed in the inset of Fig. 3. Interestingly, our calculations showed that $\mathcal{N}[\hat{\rho}_{\mathcal{S}E}(t)]$ and $Q_{\mathcal {S}E}(t)$, which were obtained from two completely different theories, are proportional to each other. This can be observed in the time-dependent profiles shown in Fig. 3.  

\begin{figure}[ht]
\centerline{\includegraphics[width=6.9cm]{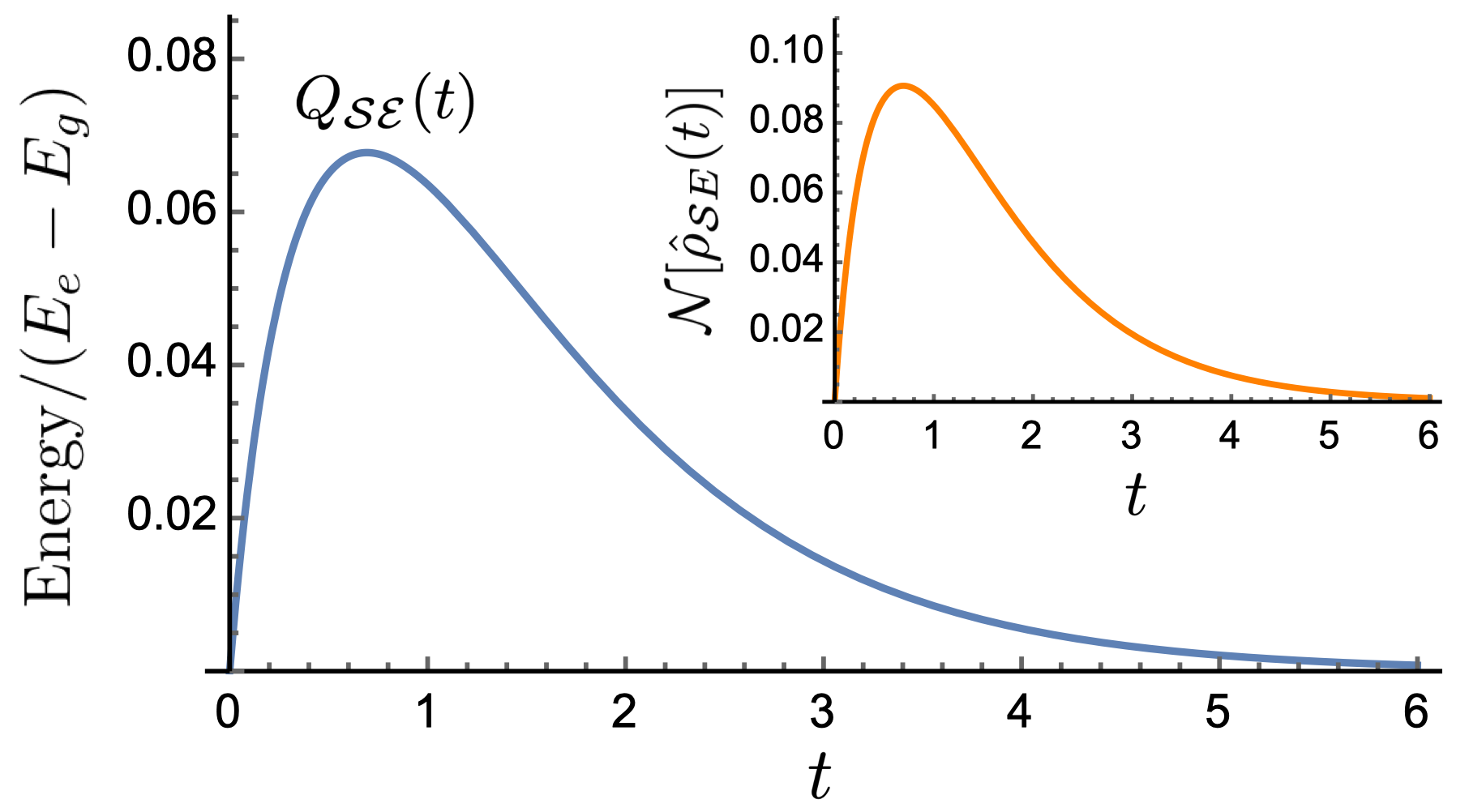}}
\caption{Time evolution of the heat asymmetry $Q_{\mathcal {S}E}(t)$ between the system and environment, which interact via the GADC (we used the same parameters of Fig. 2). If compared with the values of $Q_{\mathcal {S}}(t)$ and $Q_{E}(t)$ shown in Fig. 2, we observe that the heat asymmetry is high for short times ($t<2$). The inset presents the time evolution of the entanglement negativity $\mathcal{N}[\hat{\rho}_{\mathcal{S}E}(t)]$ created between $\mathcal {S}$ and $E$. Curiously, $Q_{\mathcal {S}E}(t)$ and $\mathcal{N}[\hat{\rho}_{\mathcal{S}E}(t)]$ are proportional in this model.} 
\end{figure}

With the results of the energy exchanges and quantum correlations between $\mathcal{S}$ and $E$ established, we can investigate the energy cost to generate entanglement in this model, which is an essential resource for many quantum information tasks \cite{nielsen}. From Eqs. (\ref{4}) to (\ref{7}) we see that the energy of the composite system is clearly conserved in the GADC, as confirmed by the result $\Delta U_{\mathcal{S}}(t) = - \Delta U_{E}(t)$. Then, we pose the question of where the energy used to create this entanglement comes from. In Fig. 2 we observe that, when the interaction begins, $E$ releases an amount of heat that is not totally absorbed by $\mathcal{S}$, and, according to Fig. 3, this is exactly when entanglement is created. After that (mostly in the interval $1 < t < 3$), $E$ still releases heat, but $\mathcal{S}$ absorbs a larger amount. This process continues until all heat released by $E$ is absorbed by $\mathcal{S}$; for long times we have $Q_{\mathcal {S}} = - Q_{E} \approx 0.104 (E_{e} - E_{g})$. In this form, we can assign the energy used to generate entanglement to the transient imbalance between the released and absorbed heat involving $\mathcal{S}$ and $E$. This justifies the proportionality between $Q_{\mathcal {S}E}(t)$ and $\mathcal{N}[\hat{\rho}_{\mathcal{S}E}(t)]$ along the entire process, and sheds a different light on the idea that entanglement is created at the cost of work \cite{huber,bruschi,beny,bera}.

\section{Conclusion}

We have presented a new framework to study the thermodynamics of an open quantum system strongly coupled to a heat bath, which takes into account the energetic aspects of quantum-mechanical resources as coherence and entanglement. The method was used to provide a thermodynamic description of the generalized amplitude-damping channel (GADC), from the point of view of both system and environment. We demonstrated that, when the interaction begins, an asymmetry between the heat released by the environment and the heat absorbed by the system emerges, while a quantum correlation is established. More specifically, it was found that the heat asymmetry in this example is proportional to the entanglement negativity during the complete time evolution. This important finding suggests that the creation of quantum correlations does not come necessarily at the price of doing work on the interacting systems. This development opens up a new venue for exploring thermodynamics at strong coupling.            
\section*{ACKNOWLEDGMENTS}
        
The author acknowledges support from Coordena{\c c}{\~a}o de Aperfei{\c c}oamento
de Pessoal de N{\'i}vel Superior (CAPES, Finance Code 001), Conselho Nacional de Desenvolvimento Cient{\'i}fico e Tecnol{\'o}gico (CNPq, Grant No. 303451/2019-0), Pronex/Fapesq-PB/CNPq, Grant No. 0016/2019, and PROPESQ/PRPG/UFPB (Project code PIA13177-2020).

\newpage

\widetext
\begin{center}
\textbf{\large Supplemental Material: Relating Heat and Entanglement in Strong Coupling Thermodynamics}
\end{center}
\setcounter{equation}{0}
\setcounter{figure}{0}
\setcounter{table}{0}
\setcounter{page}{1}
\makeatletter
\renewcommand{\theequation}{S\arabic{equation}}
\renewcommand{\thefigure}{S\arabic{figure}}
\renewcommand{\bibnumfmt}[1]{[S#1]}
\renewcommand{\citenumfont}[1]{S#1}

In this Supplemental Material we give some details about the calculations of the thermodynamic and information-theoretic properties of the system and environment presented in the main text. In particular, we introduce a discussion of the quantum coherence and quantum mutual information in the physical model studied.

\section{Thermodynamic properties of $\mathcal{S}$ and $E$}

\subsection{Heat and coherent energy}

We now discuss the thermodynamic properties of the system $\mathcal{S}$ and the qubit environment $E$ interacting via the generalized amplitude-damping channel (GADC), according to Eqs. (2) to (4), whose formalism was first derived by the author in Ref. \cite{Bert2}. As pointed out in the main text, no work is involved in the process, i.e., $W_{\mathcal{S}}(t) = W_{E}(t) = 0$. However, to calculate the heat and coherent energy we need the eigenvalues and eigenstates of the Hamiltonian and the density matrix of both $\mathcal{S}$ and $E$. The free Hamiltonian of the system is
\begin{equation}
\label{S1}
\hat{H}_{\mathcal{S}} = E_{g}\ket{g}\bra{g} + E_{e}\ket{e}\bra{e},
\end{equation}
whose eigenvalues are $E_{g}$ and $E_{e}$, and the respective eigenstates are $\ket{g}$ and $\ket{e}$. Conversely, the free Hamiltonian of $E$ is
\begin{equation}
\label{S2}
\hat{H}_{E} = E_{0}\ket{E_{0}}\bra{E_{0}} + E_{1}\ket{E_{1}}\bra{E_{1}},
\end{equation}
with eigenvalues $E_{0}$ and $E_{1}$, and  eigenstates  $\ket{E_{0}}$ and $\ket{E_{1}}$, respectively.

On the other hand, the time-dependent quantum states of $\mathcal{S}$ and $E$ are given in Eq. (10) of the main text, which can be explicitly written as 
\begin{align}
\label{S3}
\hat{\rho}_{\mathcal{S}}(t)  = 
\begin{bmatrix}
    [\alpha^2 + (1 - \alpha^2)\delta(t)]w_{0} + \alpha^2 \gamma (t) w_{1} & \alpha \sqrt{1 - \alpha^2} [\gamma (t)]^{1/2} \\
    \alpha \sqrt{1 - \alpha^2} [\gamma (t)]^{1/2} & (1 - \alpha^2)\gamma(t) w_{0} + [(1-\alpha^2) + \alpha^2 \delta(t)] w_{1}\\
\end{bmatrix}
\end{align}
and
\begin{align}
\label{S4}
\hat{\rho}_{E}(t) =
\begin{bmatrix}
     [\alpha^2 + (1 - a^2)\gamma(t)]w_{0} + \alpha^2 \delta (t) w_{1} & \alpha \sqrt{1 - \alpha^2} [\delta (t)]^{1/2} \\
    \alpha \sqrt{1 - \alpha^2} [\delta (t)]^{1/2} & (1 - \alpha^2)\delta(t) w_{0} + [(1-\alpha^2) + \alpha^2 \gamma(t)] w_{1}\\
\end{bmatrix},
\end{align} 
where $\gamma (t) = e^{ -t}$ and $\delta (t) = 1- e^{- t}$. In this case, we already considered $\Gamma =1$, which is the probability of a quantum transition event per unit time. We analyzed the case in which the system started out in the maximally coherent state $\ket{+} = 1/ \sqrt{2} (\ket{g} + \ket{e})$, which means $\alpha = 1/ \sqrt{2}$, and the qubit environment in a thermal state with inverse temperature given by $\beta = 1/ (E_{e} - E_{g})$, such that  $w_{0} \approx 0.73$ and $w_{1} \approx 0.27$. In this form, the quantum states of Eqs. (\ref{S3}) and (\ref{S4}) become
\begin{align}
\label{S5}
\hat{\rho}_{\mathcal{S}}(t)  \approx 
\begin{bmatrix}
    0.365 [1 + \delta(t)] + 0.135 \gamma (t) & 0.5 [\gamma (t)]^{1/2} \\
    0.5 [\gamma (t)]^{1/2} & 0.365 \gamma(t) + 0.135 [1 + \delta(t)]\\
\end{bmatrix}
\end{align}
and
\begin{align}
\label{S6}
\hat{\rho}_{E}(t) \approx
\begin{bmatrix}
      0.365 [1 + \gamma(t)] + 0.135 \delta (t) & 0.5 [\delta (t)]^{1/2} \\
    0.5 [\delta (t)]^{1/2} & 0.365 \delta(t) + 0.135 [1 + \gamma(t)]\\
\end{bmatrix},
\end{align}
respectively.

The eigenvalues of $\hat{\rho}_{\mathcal{S}}(t)$ can be found to be
\begin{equation}
\label{S7}
\rho_{0}(t) \approx 0.1 e^{-t} \left(5.07 e^t-\sqrt{5.45\, +14.85 e^t+5.45 e^{2 t}}\right)
\end{equation}
and
\begin{equation}
\label{S8}
\rho_{1}(t) \approx 0.1 e^{-t} \left(5.07 e^t + \sqrt{5.45\, +14.85 e^t+5.45 e^{2 t}} \right),
\end{equation}
with the respective eigenvectors
\begin{equation}
\label{S9}
\ket{k_{0}(t)} \approx \frac{0.46 e^{-0.5 t} \left(-0.43 \sqrt{5.45 +14.85 e^{t}+5.45 e^{2 t}}+e^{t}-1\right) \ket{g} +  \ket{e}}{\sqrt{\left(0.46 e^{-0.5 t} \left(-0.43 \sqrt{5.45 +14.85 e^{t}+5.45 e^{2 t}}+ e^{t}-1\right)\right)^2+1}}
\end{equation}
and
\begin{equation}
\label{S10}
\ket{k_{1}(t)} \approx \frac{0.46 e^{-0.5 t} \left(0.43 \sqrt{5.45 +14.85 e^{t}+5.45 e^{2 t}}+e^{ t}-1\right) \ket{g} + \ket{e}}{\sqrt{\left(0.46 e^{-0.5 t} \left(0.43 \sqrt{5.45 +14.85 e^{t}+5.45 e^{2 t}}+ e^{t}-1\right)\right)^2+1}}.
\end{equation}

We are now in a position to calculate the thermodynamic properties of the system. From Eq. (3) we have that the heat absorbed by the system as a function of time is
\begin{align}
\label{S11}
Q_{\mathcal{S}}(t)  &=  \sum_{n} \sum_{k} \int_{0}^{t}  E_{n} |c_{n,k}|^{2} \frac{d \rho_{k}}{dt'} dt' \nonumber \\
&= E_{g} \left[ \int_{0}^{t}  |\braket{g|k_{0}(t')}|^{2} \frac{d}{dt'} \rho_{0} (t') dt' + \int_{0}^{t}  |\braket{g|k_{1}(t')}|^{2} \frac{d}{dt'} \rho_{1} (t') dt' \right]\nonumber \\ &+E_{e} \left[ \int_{0}^{t}  |\braket{e|k_{0}(t')}|^{2} \frac{d}{dt'} \rho_{0} (t') dt' + \int_{0}^{t}  |\braket{e|k_{1}(t')}|^{2} \frac{d}{dt'} \rho_{1} (t') dt' \right].
\end{align}
In turn, from Eq. (4) the coherent energy of the system as a function of time is given by
\begin{align}
\label{S12}
\mathcal{C}_{\mathcal{S}}(t)
&= \sum_{n}\sum_{k} \int_{0}^{t} (E_{n} \rho_{k}) \frac{d}{dt'} |c_{n,k}|^{2} dt' \nonumber \\
&= E_{g} \left[ \int_{0}^{t}  \rho_{0}(t') \frac{d} {dt'} |\braket{g|k_{0}(t')}|^{2} dt' + \int_{0}^{t}  \rho_{1}(t') \frac{d} {dt'} |\braket{g|k_{1}(t')}|^{2} dt' \right] \nonumber \\ &+E_{e} \left[ \int_{0}^{t} \rho_{0}(t') \frac{d} {dt'} |\braket{e|k_{0}(t')}|^{2} dt' + \int_{0}^{t}  \rho_{1}(t') \frac{d} {dt'} |\braket{e|k_{1}(t')}|^{2} dt' \right].
\end{align}
The expressions of $Q_{\mathcal{S}}(t)$ and $\mathcal{C}_{\mathcal{S}}(t)$ can be calculated by substitution of Eqs. (\ref{S7}) to (\ref{S10}) into Eqs. (\ref{S11}) and (\ref{S12}). The results are too long to be written here, however, the time-dependent profiles are those shown in Fig. 2a in the main text. The behavior of the change in the internal energy of the system, $\Delta U_{\mathcal{S}}(t) = Q_{\mathcal{S}}(t) + \mathcal{C}_{\mathcal{S}}(t)$, is also shown.

Now we turn to the analysis of the thermodynamics of the environment. The eigenvalues of the state $\hat{\rho}_{E}(t)$ of Eq. (\ref{S6}) are
\begin{equation}
\label{S13}
\lambda_{0}(t) \approx 0.5 e^{-t} \left( e^t-\sqrt{0.21 - e^t + e^{2 t}}\right)
\end{equation}
and
\begin{equation}
\label{S14}
\lambda_{1}(t) \approx 0.5 e^{-t} \left( e^t + \sqrt{0.21 - e^t + e^{2 t}}\right),
\end{equation}
with the respective eigenvectors 

\begin{equation}
\label{S15}
\ket{l_{0}(t)} \approx \frac{-\frac{e^{-t} \left(\sqrt{0.21 - e^t+ e^{2 t}}-0.46\right)}{\sqrt{1-e^{-t}}} \ket{E_{0}} +  \ket{E_{1}}}{\sqrt{\left(-\frac{e^{-t} \left( \sqrt{0.21 - e^t+ e^{2 t}}-0.46\right)}{\sqrt{1-e^{-t}}}\right)^2+1}}
\end{equation}
and
\begin{equation}
\label{S16}
\ket{l_{1}(t)} \approx 
\frac{\frac{e^{-t} \left(\sqrt{0.21 - e^t+ e^{2 t}}+0.46\right)}{\sqrt{1-e^{-t}}} \ket{E_{0}} +  \ket{E_{1}}}{\sqrt{\left(\frac{e^{-t} \left( \sqrt{0.21 - e^t+ e^{2 t}}+0.46\right)}{\sqrt{1-e^{-t}}}\right)^2+1}}.
\end{equation}

Having this information, we can calculate the heat released by the environment $E$, 
\begin{align}
\label{S17}
Q_{E}(t)  &=  \sum_{m} \sum_{l} \int_{0}^{t}  E_{m} |c_{m,l}|^{2} \frac{d \lambda_{l}}{dt'} dt' \nonumber \\
&= E_{0} \left[ \int_{0}^{t}  |\braket{E_{0}|l_{0}(t')}|^{2} \frac{d}{dt'} \lambda_{0} (t') dt' + \int_{0}^{t}  |\braket{E_{0}|l_{1}(t')}|^{2} \frac{d}{dt'} \lambda_{1} (t') dt' \right]\nonumber \\ &+E_{1} \left[ \int_{0}^{t}  |\braket{E_{1}|l_{0}(t')}|^{2} \frac{d}{dt'} \lambda_{0} (t') dt' + \int_{0}^{t}  |\braket{E_{1}|l_{1}(t')}|^{2} \frac{d}{dt'} \lambda_{1} (t') dt' \right],
\end{align}
as well as the coherent energy absorbed,
\begin{align}
\label{S18}
\mathcal{C}_{E}(t)
&= \sum_{m}\sum_{l} \int_{0}^{t} (E_{m} \lambda_{l}) \frac{d}{dt'} |c_{m,l}|^{2} dt' \nonumber \\
&= E_{0} \left[ \int_{0}^{t}  \lambda_{0}(t') \frac{d} {dt'} |\braket{E_{0}|l_{0}(t')}|^{2} dt' + \int_{0}^{t}  \lambda_{1}(t') \frac{d} {dt'} |\braket{E_{0}|l_{1}(t')}|^{2} dt' \right] \nonumber \\ &+E_{1} \left[ \int_{0}^{t} \lambda_{0}(t') \frac{d} {dt'} |\braket{E_{1}|l_{0}(t')}|^{2} dt' + \int_{0}^{t}  \lambda_{1}(t') \frac{d} {dt'} |\braket{E_{1}|l_{1}(t')}|^{2} dt' \right].
\end{align}
In the end, by substitution of the results of  Eqs. (\ref{S13}) to (\ref{S16}) into Eqs. (\ref{S17}) and (\ref{S18}), we can obtain the expressions of $Q_{E}(t)$ and $\mathcal{C}_{E}(t)$. As in the case of the thermodynamic properties of the system, these expressions are also too long to be displayed. Meanwhile, the time-dependent behaviors are illustrated in Fig. 2b of the main text. The time dependence of the change in the internal energy of $E$, $\Delta U_{E}(t) = Q_{E}(t) + \mathcal{C}_{E}(t)$, is also illustrated.

\section{ Information-theoretic properties of $\mathcal{S}$ and $E$}

\subsection{Coherence, entropy and mutual information}

Now we investigate some important information-theoretic properties of the system and the environment that helps us to understand the thermodynamics of the interaction described in the main text. The first property to be studied is the coherence, which will be quantified here with the so-called {\it norm of coherence} \cite{Baum}. For a general state $\hat{\rho}$, it is given simply by $C(\hat{\rho}) = \sum_{i \neq j} |\rho_{ij}|$. Therefore, by using the results of Eqs. (\ref{S5}) and (\ref{S6}), it is easy to see that the coherence of $\mathcal{S}$ and $E$ are given respectively by $C[\hat{\rho}_{\mathcal{S}}(t)] = [\gamma (t)]^{1/2} =  e^{-t/2}$ and $C[\hat{\rho}_{E}(t)] = [\delta (t)]^{1/2} = \sqrt{1- e^{-t}}$. These results are plotted in Fig. S1(a), which shows that for short times ($t<2$) the coherence of the system is decreasing at a rate much lower than the rate of increase of the environment coherence. For longer times, this tendency is inverted. This behavior at least in part explains the fact that for short times the absorption of coherent energy by $E$ is faster than the release of coherent energy by $\mathcal{S}$, as can be observed in Fig. 2 in the main text.

Another essential property to be addressed here is the von Neumann entropy of both $\mathcal{S}$ and $E$. They are given by $S[\hat{\rho}_{\mathcal{S}}(t)] = -$tr$\{\hat{\rho}_{\mathcal{S}}(t) \log \hat{\rho}_{\mathcal{S}}(t)\} =  - \sum_{k = 0,1} \rho_{k}(t) \log \rho_{k}(t)$ and $S[\hat{\rho}_{E}(t)] = -$tr$\{\hat{\rho}_{E}(t) \log \hat{\rho}_{E}(t)\} =  - \sum_{l = 0,1} \lambda_{l}(t) \log \lambda_{l}(t)$. The logarithm
is base 2, so that $S$ is measured in qubits. Using the results of Eqs. (\ref{S7}), (\ref{S8}), (\ref{S13}) and (\ref{S14}), we can calculate the entropies, whose time-dependent behaviors are displayed in Fig. S1(b). Note that the entropy increase of the system and the entropy decrease of the environment clearly occur at different rates for short times. To some extent, this fact is related to the manifestation of the heat asymmetry pointed out in the main text, once heat was defined as: ``the change in the internal energy that is accompanied by entropy change''.

\begin{figure}[ht]
\centerline{\includegraphics[width=15.5cm]{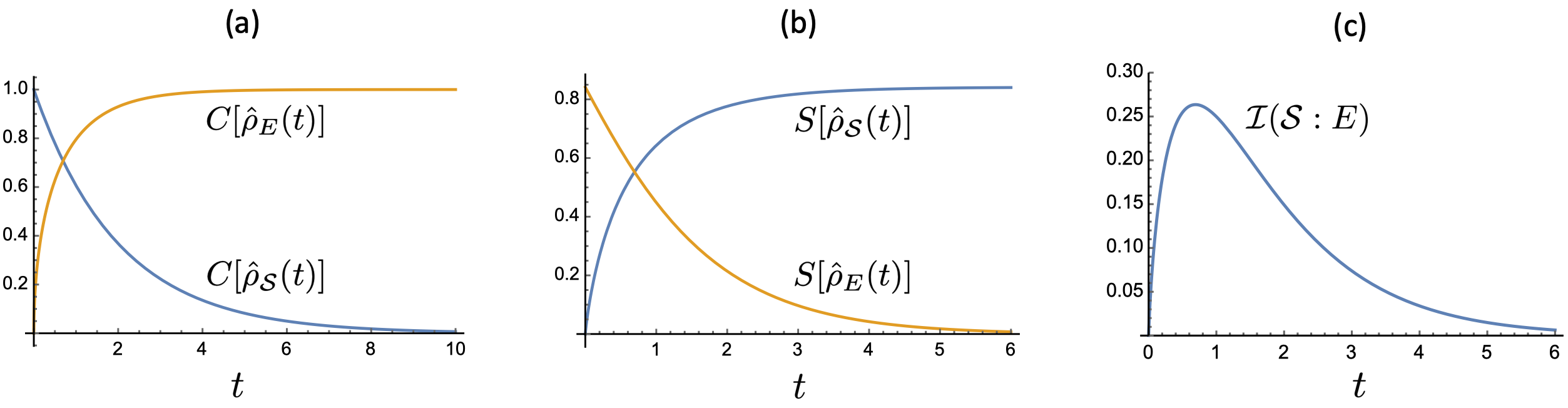}}
\caption{Time-dependence of: (a) the norm of coherence and (b) the von Neumann entropy of $\mathcal{S}$ and $E$. The behavior of the quantum mutual information is shown in panel (c).}
\end{figure}

In the present context, it is also relevant to study the time evolution of the quantum mutual information between $\mathcal{S}$ and $E$, which is an alternative form of measuring quantum correlations \cite{Nielsen,Preskill}. This is given by

\begin{equation}
\label{S19}
\mathcal{I}(\mathcal{S}:E)(t) = S[\hat{\rho}_{\mathcal{S}}(t)] + S[\hat{\rho}_{E}(t)] - S[\hat{\rho}_{\mathcal{S}E}(t)],
\end{equation}
where $S[\hat{\rho}_{\mathcal{S}E}(t)]$ is the joint entropy of $\mathcal{S}$ and $E$. The composite system evolves under the probabilistic unitary transformation $\hat{\mathcal{U}}$ given in Eq. (9) of the main text. However, since the von Neumann entropy is invariant under unitary transformations, we have that $S[\hat{\rho}_{\mathcal{S}E}(t)] = S[\hat{\mathcal{U}}\hat{\rho}_{\mathcal{S}E}(0)\hat{\mathcal{U}}^{\dagger}] = S[\hat{\rho}_{\mathcal{S}E}(0)] = S[\hat{\rho}_{\mathcal{S}}(0) \otimes \hat{\rho}_{E}(0)]$. We also have that the von Neumann entropy is additive for uncorrelated systems, i.e., $S[\hat{\rho}_{\mathcal{S}}(0) \otimes \hat{\rho}_{E}(0)] = S[\hat{\rho}_{\mathcal{S}}(0)]+ S[ \hat{\rho}_{E}(0)] = S[ \hat{\rho}_{E}(0)] $. In the last equality we used the fact that $S[\hat{\rho}_{\mathcal{S}}(0)] = 0$, because $\hat{\rho}_{\mathcal{S}}(0)$ is a pure state. With this, we have that $S[\hat{\rho}_{\mathcal{S}E}(t)] = S[ \hat{\rho}_{E}(0)]  =  - \sum_{l = 0,1} \lambda_{l}(0) \log \lambda_{l}(0) \approx 0.84$. This allows us to calculate the time evolution of the quantum mutual information according to Eq. (\ref{S19}), whose behavior is shown in Fig. S1(c). Remarkably, the profile of $\mathcal{I}(\mathcal{S}:E)(t)$ is very similar to those of the
heat asymmetry $Q_{\mathcal {S}E}(t)$ and the entanglement negativity $\mathcal{N}[\hat{\rho}_{\mathcal{S}E}(t)]$ shown in Fig. 3 of the main text. Nevertheless, it can be verified that only $Q_{\mathcal {S}E}(t)$ and $\mathcal{N}[\hat{\rho}_{\mathcal{S}E}(t)]$ are proportional to each other.

\subsection{Entanglement negativity}

Now we detail the calculation of the entanglement negativity discussed in the main manuscript. To begin with, we express the matrix of the initial uncorrelated state in the basis $\{ \ket{g,E_{0}},\ket{g,E_{1}},\ket{e,E_{0}},\ket{e,E_{1}} \}$:    
\begin{align}
\label{S20}
\hat{\rho}_{\mathcal{S}E}(0) = \hat{\rho}_{\mathcal{S}}(0) \otimes \hat{\rho}_{E}(0) = 
\begin{bmatrix}
      \alpha^2 w_{0} & 0 & \alpha \sqrt{1 - \alpha^2} w_{0} & 0 \\
    0 & \alpha^2 w_{1} & 0 & \alpha \sqrt{1 - \alpha^2} w_{1} \\  \alpha \sqrt{1 - \alpha^2} w_{0} & 0 & (1 - \alpha^2) w_{0} & 0 \\
    0 & \alpha \sqrt{1 - \alpha^2} w_{1} & 0 & (1 - \alpha^2) w_{1} 
\end{bmatrix},
\end{align}
where we are still considering that the system started out in the general pure state $\ket{\psi(0)} = \alpha \ket{g} + \sqrt{1 - \alpha^2} \ket{e}$, and the environment qubit in the state $\hat{\rho}_{E}(0) = w_{0}\ket{E_{0}}\bra{E_{0}} + w_{1}\ket{E_{1}}\bra{E_{1}}$. The evolution of the composite system is obtained by applying the probabilistic unitary transformation given in Eq. (9) of the main text, with the probability rules considered, $p \rightarrow 1 - e
^{- t}$. Therefore, we have 

\begin{align}
\label{S21}
\nonumber
 &\hat{\rho}_{\mathcal{S}E}(t) = \hat{\mathcal{U}} \hat{\rho}_{\mathcal{S}E}(0)  \hat{\mathcal{U}}^{\dagger}\\ 
&= 
\begin{bmatrix}
      \alpha^2 w_{0} & \alpha \sqrt{1 - \alpha^2} w_{0} [\delta (t)]^{1/2} & \alpha \sqrt{1 - \alpha^2} w_{0} [\gamma (t)]^{1/2}  & 0 \\
    \alpha \sqrt{1 - \alpha^2} w_{0} [\delta (t)]^{1/2} & \alpha^2 w_{1} \gamma (t) + (1 - \alpha^2) w_{0} \delta (t) & [\alpha^2 w_{1} + (1 - \alpha^2) w_{0}] [ \delta(t) \gamma (t)]^{1/2} & \alpha \sqrt{1 - \alpha^2} w_{1} [\gamma(t)]^{1/2} \\  \alpha \sqrt{1 - \alpha^2} w_{0} [\gamma(t)]^{1/2} & [\alpha^2 w_{1} + (1 - \alpha^2) w_{0}] [ \delta(t) \gamma (t)]^{1/2} &  \alpha^2 w_{1} \delta (t) + (1 - \alpha^2) w_{0} \gamma (t) & \alpha \sqrt{1 - \alpha^2} w_{1} [\delta(t)]^{1/2} \\
     0 & \alpha \sqrt{1 - \alpha^2} w_{1} [\gamma(t)]^{1/2} & \alpha \sqrt{1 - \alpha^2} w_{1} [\delta(t)]^{1/2} & (1 - \alpha^2) w_{1} 
\end{bmatrix}.
\end{align}
In turn, the partial transpose of $\hat{\rho}_{\mathcal{S}E}(t)$, with the parameters fixed in the main text ($\alpha = 1/ \sqrt{2}$, $w_{0} \approx 0.73$ and $w_{1} \approx 0.27$), is given by  
\begin{align}
\label{S22}
\hat{\rho}_{\mathcal{S}E}^{T_{\mathcal{S}}}(t) = 
 \begin{bmatrix}
      0.365 & 0.365 [\delta (t)]^{1/2} & 0.365 [\gamma (t)]^{1/2}  & 0.5 [ \delta(t) \gamma (t)]^{1/2} \\
    0.365 [\delta (t)]^{1/2} & 0.135 \gamma (t) + 0.365 \delta (t) & 0 & 0.135 [\gamma(t)]^{1/2} \\  0.365 [\gamma(t)]^{1/2} & 0 &  0.135 \delta (t) + 0.365 \gamma (t) & 0.135 [\delta(t)]^{1/2} \\
    0.5 [ \delta(t) \gamma (t)]^{1/2} & 0.135[\gamma(t)]^{1/2} & 0.135 [\delta(t)]^{1/2} & 0.135 
\end{bmatrix}.
\end{align}
At this point, the time-dependence of the entanglement negativity can be obtained as the sum of the absolute values of the negative eigenvalues of $\hat{\rho}_{\mathcal{S}E}^{T_{\mathcal{S}}}(t)$. The result can be calculated numerically, but it is too cumbersome to be written here. However, the time-dependent behavior is shown in the inset of Fig. 3 of the main text.

\end{document}